\documentclass[aps,twocolumn,showpacs,preprintnumbers,prl,amsmath,amssymb,amsfonts,superscriptaddress,floatfix]{revtex4-1}

\usepackage[paperwidth=210mm,paperheight=297mm,centering,hmargin=2.3cm,tmargin=2.8cm,bmargin=3.4cm]{geometry}
\usepackage{booktabs}
\usepackage{subfigure}
\usepackage{graphicx}
\usepackage{amsfonts}
\usepackage{amssymb}
\usepackage{amsmath}
\usepackage{overpic}
\usepackage{braket}
\usepackage{xcolor}
\usepackage{cancel}
\usepackage{fancyhdr}
\usepackage[normalem]{ulem}
\usepackage{dblfloatfix}

\usepackage{bm}
\usepackage{epstopdf}

\newcommand{\be}{\begin{equation}}
\newcommand{\ee}{\end{equation}}
\newcommand{\bea}{\begin{eqnarray}}
\newcommand{\eea}{\end{eqnarray}}

\usepackage{xcolor}
\definecolor{highlight}{rgb}{0, 0.38, 1}
\definecolor{obs}{rgb}{1, 0, 1}
\definecolor{todo}{rgb}{1, 0, 0}

\def\lb{\label}
\def\pref#1{(\ref{#1})}

\newcount\bozza \bozza=1
\ifnum\bozza=1
\newdimen\shift \shift=-2truecm

\allowdisplaybreaks

\makeatletter
\makeatletter \renewcommand{\fnum@figure}{{\bf{\figurename~\thefigure}}}
\makeatother

\begin{document}

\title{Transport signatures of fragile-glass dynamics \\ in the melting of the two-dimensional vortex lattice}
\author{I. Maccari{$^{\dag}$}}
\affiliation{Department of Physics, Stockholm University, 106 91 Stockholm, Sweden}
\author{Bal~K.~Pokharel{$^{\dag}$}}
\affiliation{National High Magnetic Field Laboratory, Florida State University, Tallahassee, FL 32310, USA}
\affiliation{Department of Physics, Florida State University, Tallahassee, Florida 32306, USA}
\author{J. Terzic} 
\affiliation{National High Magnetic Field Laboratory, Florida State University, Tallahassee, FL 32310, USA}
\affiliation{Department of Physics and Astronomy, Western Kentucky University, Bowling Green, KY 42101, USA}
\author{Surajit Dutta}
\affiliation{Tata Institute of Fundamental Research, Homi Bhabha Rd, Colaba, Mumbai 400005, India}
\author{J. Jesudasan}
\affiliation{Tata Institute of Fundamental Research, Homi Bhabha Rd, Colaba, Mumbai 400005, India}
\author{Pratap Raychaudhuri}
\affiliation{Tata Institute of Fundamental Research, Homi Bhabha Rd, Colaba, Mumbai 400005, India}
\author{J. Lorenzana}
\affiliation{Institute for Complex Systems (ISC-CNR), UOS Sapienza, P.le A. Moro 5, 00185 Rome, Italy}

\author{C. De Michele}
\affiliation{Department of Physics, Sapienza University of Rome, P.le A. Moro 2, 00185 Rome, Italy}

\author{C. Castellani}
\affiliation{Department of Physics, Sapienza University of Rome, P.le A. Moro 2, 00185 Rome, Italy}
\affiliation{Institute for Complex Systems (ISC-CNR), UOS Sapienza, P.le A. Moro 5, 00185 Rome, Italy}

\author{L.Benfatto{$^{\ast}$}}
\affiliation{Department of Physics, Sapienza University of Rome, P.le A. Moro 2, 00185 Rome, Italy}
\affiliation{Institute for Complex Systems (ISC-CNR), UOS Sapienza, P.le A. Moro 5, 00185 Rome, Italy}

\author{{Dragana~Popovi\'c}{$^{\ast}$}}
\affiliation{National High Magnetic Field Laboratory, Florida State University, Tallahassee, FL 32310, USA}
\affiliation{Department of Physics, Florida State University, Tallahassee, Florida 32306, USA}

\maketitle
\noindent

\noindent {\normalsize{$^\dag$ \textup{These authors contributed equally to this work.}}}\linebreak
{\normalsize{$^\ast$~To whom correspondence should be addressed; Email: lara.benfatto@roma1.infn.it, }\\
{\normalsize {dragana@magnet.fsu.edu}}

\vspace{0.5 cm}

\textbf{In two-dimensional (2D) systems, the melting from a solid to an isotropic liquid can occur via an intermediate phase that retains orientational order.  However, in 2D superconducting vortex lattices, the effect of orientational correlations on transport, and their interplay with disorder remain open questions.  Here we study a 2D weakly pinned vortex system in amorphous MoGe films over an extensive range of temperatures ($\bm{T}$) and perpendicular magnetic fields ($\bm{H}$) using linear and nonlinear transport measurements.  We find that, at low fields, the resistivity obeys the Vogel-Fulcher-Tamman (VFT) form, $\bm{\rho(T)\propto\exp[-{W}(H)/(T-T_0(H))]}$, characteristic of fragile glasses.  As $\bm{H}$ increases, $\bm{T_0(H)}$ is suppressed to zero, and a standard vortex liquid behavior consistent with a $\bm{T=0}$ superconducting transition is observed.  Our findings, supported also by simulations, suggest that the presence of orientational correlations gives rise to a heterogeneous dynamics responsible for the VFT behavior.  The effects of quenched disorder become dominant at high $\bm{H}$, where a crossover to a strong-glass behavior is observed.  This is a new insight into the dynamics of melting in 2D systems with competing orders.}

Thermal melting of 2D crystalline solids has been investigated in a variety of systems, including colloids, electrons on the surface of liquid helium, rare-gas atoms on substrates such as graphite, liquid crystal films, dust plasmas, and vortices in thin superconducting (SC) 
films in a magnetic field \cite{kosterlitz_kosterlitzthouless_2016, ryzhov_berezinskiikosterlitzthouless_2017}.  The melting transition is generally understood to be driven by the proliferation of topological defects, as described by the Berezinskii-Kosterlitz-Thouless-Halperin-Nelson-Young (BKTHNY) theory \cite{halperin_1978, nelson_1979, Young_1979}.  According to BKTHNY, for weak enough quenched disorder the transition from the 2D (weakly pinned) solid to the liquid phase occurs via an intermediate phase called hexatic. In the hexatic phase, free dislocations appear, breaking the quasi-long-range positional order, but preserving the hexagonal orientational one.  By increasing $T$ further, free disclinations form and a fully isotropic liquid is established.  This melting sequence has indeed been detected by scanning-tunneling-spectroscopy (STS) imaging of vortices in 2D superconductors \cite{guillamon_direct_2009,guillamon_enhancement_2014, Roy_2019, Dutta_2019}, but the morphology of dislocations in real systems may be more complicated and depend on microscopic details \cite{kosterlitz_kosterlitzthouless_2016, ryzhov_berezinskiikosterlitzthouless_2017}.  
For example, an STS study of W-based SC films reported a coexistence of liquid with hexatic, as well as with smectic-like (striped) regions of vortex arrangements, and suggested that this might be a feature of melting of 2D solids formed by linear-like units, including vortices in Abrikosov lattices \cite{guillamon_direct_2009}.  For weak pinning and intermediate $H$, numerical studies also support \cite{Zimany_PRB2004} the appearance of grain boundaries, formed by chains of dislocations, which separate the domains of topologically ordered vortex lattice.  The presence of competing orders may give rise to metastable states and the associated slow dynamics in many systems \cite{Dagotto2005}.  An STS study on amorphous MoGe (\textit{a}-MoGe) thin films did observe \cite{Dutta_2019} a strong suppression of the vortex diffusivity in the presence of hexatic correlations compared to the isotropic liquid, but the nature of the dynamics near the melting transition and its effect on electrical transport have not been explored.  Another open question is the evolution of transport properties with $H$ in the presence of orientational correlations. The random potential generated by quenched impurities is indeed coupled to the vortex density, so that the increase of $H$ increases the effective disorder \cite{Giamarchi_2009}.  
 
Here, we perform extensive magnetotransport measurements on \textit{a}-MoGe thin films similar to those used in the STS studies that revealed the presence of hexatic correlations at low $H$ \cite{Roy_2019, Dutta_2019}.  Our key experimental finding is the vanishing of the resistivity, in the VFT fashion, at a nonzero temperature $T_0(H)$, for fixed low fields $H<9$~T.  The VFT law \cite{vogel_1921, fulcher_1925, tamman_1926} generally describes the behavior of the so-called fragile liquids above the glass transition, and it is usually attributed to the emergence of dynamical heterogeneities  \cite{CAVAGNA200951}.   As $H$ increases, we find that $T_0(H)$ is suppressed to zero and remains zero over a finite range of $H$ ($9<H\lesssim12.5$~T).  In this high-field range, both the observed Arrhenius dependence ($T_0=0$) of the linear resistivity and nonlinear transport (i.e. voltage-current characteristics $V$--$I$) are consistent with the disorder-induced freezing of an isotropic vortex liquid to an amorphous vortex glass phase at $T=0$.  Our finding of the fragile-glass behavior at low $H$, i.e. in the regime where disorder is not dominant, strongly suggests that the slow vortex dynamics described by the VFT law results from dynamical heterogeneities induced by emergent orientational correlations as the melting transition is approached.  To elucidate this connection further, we complement our experimental work with Monte Carlo simulations of the 2D XY model in the presence of a transverse magnetic field, analyzing both the static and the dynamic properties of the system.  We find that the orientational order leads to a caging effect that suppresses vortex diffusivity, analogous to caging effects induced by disorder in fragile glass-forming liquids that exhibit VFT behavior \cite{CAVAGNA200951}.  Indeed, we identify several signatures of the heterogeneous nature of the vortex dynamics in the presence of hexatic correlations.  Our study of transport thus reveals a fragile-glass-like dynamics associated with the thermal melting of a weakly pinned  2D vortex lattice, and a crossover to a strong-glass ($T_0=0$) behavior at higher $H$, resulting from the interplay of orientational correlations and disorder.
\onecolumngrid

\begin{figure}[b!] 
\includegraphics[width=0.948\linewidth,clip=]{{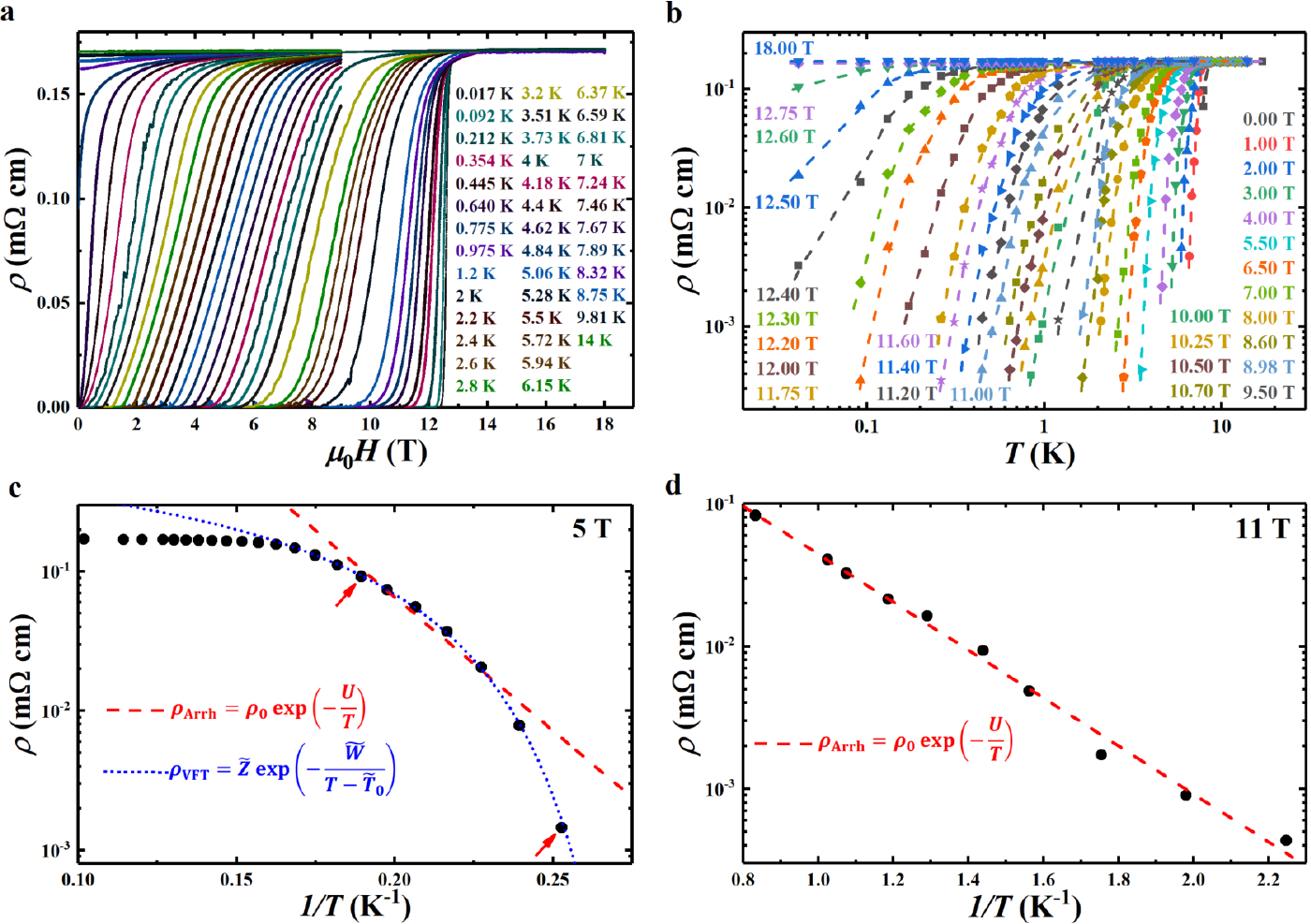}}
\caption{ \label{Figdata} {\bf Resistivity of a 22 nm thick $\bm{a}$-MoGe film at different $\bm{T}$ and $\bm{H}$.}  {The data are shown for sample 1.}  \textbf{a}, $\rho$ vs $H$ for several $T$, ranging from 0.017~K to 14~K. \textbf{b}, $\rho(T)$ at selected $H$ extracted from the data in \textbf{a}.  The dashed lines guide the eye. \textbf{c}, $\rho(T)$ fitted with the Arrhenius (red dashed line) and VFT (blue dotted line) forms for $H=5$~T; the fits are performed for the data points within the two red arrows.  $\rho$ at the lowest $T$ drops much faster than predicted by the Arrhenius function, but such a fast drop is captured well by the VFT fit; here $\tilde T_0=(3.42\pm 0.01)$~K [the fit to Eq.\ \pref{Dv} gives $T_0=(3.37 \pm 0.01)$~K].  In fact, the VFT fit describes the data even at somewhat higher $T$, beyond the red arrow.  \textbf{d}, For $H=11$~T, all the low-$T$ data are described by the Arrhenius form (or VFT form with $T_0 \approx 0$~K), as shown by the red dashed curve.  
     }
\end{figure}
\clearpage

\twocolumngrid

\noindent\textbf{Linear resistivity measurements}\\
Our samples are 22~nm thick $a$-MoGe films with SC transition temperatures $T_\mathrm{c}\sim 7.6$~K at zero field (see Methods), and low normal-state sheet resistances $R_\mathrm{s}\approx78~\Omega$ indicative of a very weak disorder \cite{Graybeal1984, Dutta_2019}.  Here $T_\mathrm{c}$ is defined as the temperature at which the linear resistance $R\equiv\lim_{I\rightarrow 0} V/I$ (i.e. $\rho$) becomes zero, that is, falls below the experimental noise floor (see also Methods).  The characteristic length scales for vortex distortions parallel to the applied $H$ and caused by thermal fluctuations or pinning are of the order of several $\mu$m \cite{dutta2019collective,yazdani_competition_1994}, i.e. much larger than the sample thickness.   Therefore, the vortex lattice (VL) in these films is indeed 2D, although the SC state is 3D (thickness $>\xi$, where $\xi\sim 5$~nm is the SC coherence length \cite{dutta2019collective}). 

Measurements of $\rho(H)$ at fixed $T$ (Fig.~\ref{Figdata}a) were used to determine $T_\mathrm{c}(H)$ (i.e. the corresponding field $H_\mathrm{c}$ for a given $T$) and the upper critical field $H_{\mathrm{c}2}(T)$, which we define as the magnetic field where $\rho=0.95\,\rho_\mathrm{N}$ at a given $T$ ($\rho_\mathrm{N}=0.17$~m$\Omega$\,cm is the normal-state resistivity).  We note that, since $T_\mathrm{c}(H)$ depends on the experimental resolution, it does not necessarily allow one to determine the real boundary between the liquid and solid phases, nor the possible transition to an orientational liquid phase.  

To explore the melting of the VL, we extract $\rho(T)$ curves at fixed $H$  (Fig.~\ref{Figdata}b; also Fig.~\ref{RT-linear}a).  A rapid, orders-of-magnitude change of $\rho$ with $T$, observed for fields below about 12~T and with no sign of saturation at low $T$, suggests an exponential $\rho(T)$ dependence.   Indeed, the Arrhenius law, 
\bea
\lb{Arr}
\rho_{\mathrm{Arrh}}&= \rho_0 \exp(-\frac{U(H)}{T}), 
\eea
is often used to describe the linear resistivity in the vortex liquid regime at low enough $T$ \cite{Blatter1994}.  In this, thermally assisted flux flow (TAFF) picture, vortices move collectively and overcome pinning barriers via thermal excitations when $T \lesssim U(H)$.  At the same time, although $\rho$ is nonzero, $V$-$I$ remains non-ohmic for $I\neq 0$ \cite{Giamarchi_2009}.  We note that Eq.\ \pref{Arr} assumes that $\rho$ is finite at all nonzero $T$.  However, when the superfluid phase sets in, which corresponds to a solid phase for the VL, $\rho$ vanishes, so deviations from the behavior \pref{Arr} should be observed whenever the critical temperature is finite and the transition is not strongly first order. 

We have performed fits of the $\rho(T)$ data, for fixed $H$, according to Eq.\  \pref{Arr}.   Two examples are shown in Fig.~\ref{Figdata}, one for low fields ($H=5$~T, panel c) and another for high fields ($H=11$~T, panel d).  While at high $H$ the Arrhenius fit works extremely well down to $T_\mathrm{c}$, at low fields we systematically find significant deviations (see also Fig.~\ref{Exp-vs-VFT}), with a faster suppression of $\rho$ than predicted by Eq.\ \pref{Arr}.  Interestingly, such deviation can be captured very well by the VFT law, 
\bea
\lb{VFT}
\rho_{\mathrm{VFT}}&= \tilde Z \exp(-\frac{\tilde W(H)}{T-{\tilde {T_0}(H)}}).
\eea
Here $\tilde W(H)$ is independent of $T$, and $\tilde{T_0}(H)$ provides the temperature where $\rho$ is expected to vanish.  Since we are interested in the vortex diffusivity, we have also performed fits to $\rho(T)=(h/2e)^{2}n_{\mathrm{v}}D_{\mathrm{v}}/(k_{B}T)$, where $n_\mathrm{v}=B/\Phi_0$ is the vortex density ($\Phi_0=h/2e$ is the quantum of magnetic flux attached to a single vortex, $h$ is Planck's constant, $e$ is electron charge), $k_\mathrm{B}$ is the Boltzmann constant, and the vortex diffusion coefficient 
\bea
\lb{Dv}
D_\mathrm{v}=Z \exp[-W(H)/(T-T_{0}(H)].
\eea 
For both \pref{VFT} and \pref{Dv}, the fitting parameters were extracted using global minimization, and typically found to be the same within error (see, e.g., Fig.~\ref{Figdata}c and Fig.~\ref{Exp-vs-VFT}), that is, $\tilde Z$ and $Z$ are the same within the proportionality constant between $\rho$ and $D_v$; this is consistent with the exponential term dominating $\rho(T)$.  Hence, hereafter we present only the VFT fitting parameters obtained using Eq.\ \pref{Dv}.  

The results for $T{_0}(H)$ are shown in Fig.\ \ref{fig:sketchy_phasediag}, along with the values of $T_\mathrm{c}(H)$ and $H_{\mathrm{c}2}(T)$ (see also Fig.~\ref{RT-linear}b;  Fig.~\ref{VFTFitparadep} shows all fitting parameters).
\begin{figure}[!tb]
    \centering
             \includegraphics[width=\linewidth]{{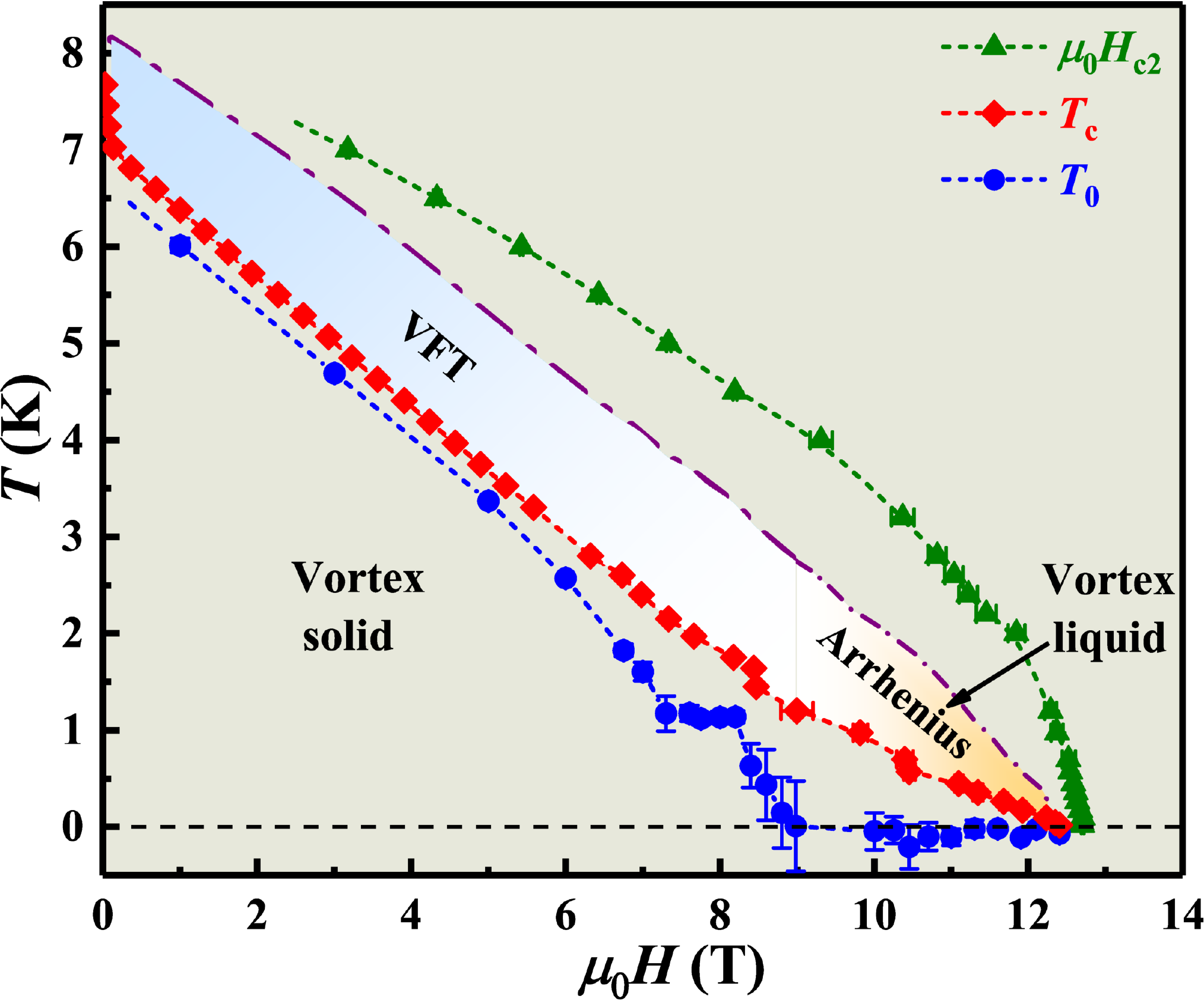}}
    \caption{ \label{fig:sketchy_phasediag} {\bf $\bm{(T,H)}$ phase diagram of the 22 nm thick $\bm{a}$-MoGe film}.  The data are shown for sample 1.  $T_{0}(H)$ (blue dots) are obtained from the VFT fits; the error bars reflect the standard errors of the fits.  In the shaded blue ``VFT'' region, $\rho(T)$ obeys the VFT law for a fixed $H\lesssim 9$~T.  For $H\gtrsim 9$~T, the resistivity above $T_\mathrm{c}$, in the shaded orange ``Arrhenius'' region, can be only fitted with the Arrhenius law \pref{Arr}.  The purple dot-dashed line shows the high-$T$ extent of the VFT and Arrhenius fits.  Green triangles: $H_{\mathrm{c}2}(T)$, i.e. the upper critical field; here $\rho$ reaches 95$\%$ of its normal-state value.  Red diamonds:  $T_\mathrm{c}(H)$, where $\rho$ drops below the experimental noise floor.  Error bars in $T_\mathrm{c} (H)$ and $H_{\mathrm{c}2}$ represent the uncertainty in defining the magnetic field corresponding to each quantity within our experimental resolution.
    }
\end{figure}
We note that, when comparing the  Arrhenius and the VFT fits, we paid attention not only to the quality of the fit, but also to the range of temperatures where either one was effective; those ranges are also shown in Fig.\ \ref{fig:sketchy_phasediag}.  As a result of such analysis, we find that for fields below $H^{\ast}\simeq 9$~T the VFT fit gives a significantly better description of the data.   With increasing $H$, however, $T_0$ decreases gradually, and vanishes near $H^{\ast}\simeq 9$~T.  For $9\lesssim H \lesssim 12.5$~T, $\rho(T)$ curves are well described by the Arrhenius fits, consistent with $T_0=0$ (Fig.~\ref{VFTB-expU}).  Furthermore, in this regime, we find that $U(H)=U_{0}\ln(H_0/H)$ [Fig.~\ref{Expo-highFields}; ${U_0}= (31.5 \pm 0.4)$~K and ${H_0}=(12.4 \pm 0.5)$~T], as expected from the TAFF model and the logarithmic vortex-vortex interactions in 2D \cite{Blatter1994}.  \\
  
\noindent\textbf{Phase diagram}\\
The behavior observed in the high-field ($H>H^{\ast}$, $T_0=0$, $T>T_\mathrm{c}$) regime (``Arrhenius'' in Fig.\ \ref{fig:sketchy_phasediag}) is, therefore, consistent with the thermally-activated collective motion of vortices in the presence of strong pinning, similar to previous studies of disordered SC films, including $a$-MoGe \cite{Ephron1996,ienaga2020quantum}.  This conclusion is further supported by our measurements of the differential resistance ($dV/dI$) versus dc current bias $I_{\mathrm{dc}}$ at fixed $H$ (see Methods).  Indeed, here the $V$--$I$ characteristic at low enough $T$ remains non-ohmic for $I_{\mathrm{dc}}\neq 0$ and $dV/dI$ increases with $I_{\mathrm{dc}}$ (Fig.~\ref{dVdI}a,~b).  Such nonlinear transport is expected from the motion of vortices in the presence of disorder, i.e. it is a signature of a viscous vortex liquid \cite{Giamarchi_2009}.  As $T$ increases, nonlinear behavior is no longer observed (Fig.~\ref{dVdI}c,~d).  At $T<T_\mathrm{c}$, where $\rho$ drops below our noise floor, a finite $I_{\mathrm{dc}}$ (i.e. a critical current) is then needed to depin a measurable number of vortices (Fig.~\ref{dVdI}e).  Since $\rho(T)$ in the liquid phase can only be fitted with the Arrhenius law \pref{Arr}, our transport results suggest that in the high-field regime an isotropic vortex liquid freezes into an amorphous vortex glass as $T\rightarrow 0$, i.e. that the solid phase is only realized at $T=0$.

In the normal state ($H>H_{\mathrm{c}2}$), $V$--$I$ characteristics are ohmic (Fig.~\ref{dVdI}b, e), as expected.  In the low-field ($H<H^{\ast}$, $T_0\neq 0$, $T>T_\mathrm{c}$) regime (``VFT'' in Fig.\ \ref{fig:sketchy_phasediag}) we also find Ohmic behavior (Fig.~\ref{dVdI}f).  However, since this regime is measured only at relatively high $T$, in analogy with the Arrhenius region we speculate that any nonlinear transport that might be present for small $I_{\mathrm{dc}}$ at lower $T$ is experimentally inaccessible. Nevertheless, this leads us to the question about the origin of the observed VFT behavior.

Within the context of glasses, one can define a temperature $T_0$, where the relaxation time diverges and the configurational entropy vanishes~\cite{CAVAGNA200951}.  $T_0$ is below the temperature $T_\mathrm{g}$ where the dynamical glass transition occurs, the value of which depends on the sensitivity of the probe. The VFT fit allows one to overcome such an experimental upper bound and to infer $T_0$, that is, the temperature scale at which the residual entropy is comparable to that of the ordered state.  Within the context of our experiment, the role of $T_\mathrm{g}$ is then played by $T_\mathrm{c}$, where the resistance drops below the measurable threshold, while $T_0$, where the diffusion coefficient (or $\rho$) vanishes, corresponds to the temperature where the transition to the truly superfluid phase occurs.  In that case, we expect that the ``VFT'' region above $T_\mathrm{c}$ shows the extent of the dynamically heterogeneous vortex liquid in the $(T, H)$ phase diagram (Fig.~\ref{fig:sketchy_phasediag}).  Interestingly, analogies to glasses were used to propose the VFT law \pref{VFT} to describe the slowing down and freezing of the strongly disordered 3D vortex matter \cite{reichhardt2000vortices}.   {However,} in our $a$-MoGe films the disorder-dominated high-field regime ($T_0=0$) is described, in contrast, by the TAFF model \pref{Arr}.  Thus our results strongly suggest that disorder is not the main origin of the VFT suppression of the vortex diffusivity observed at lower fields.   To explore the possibility that the presence of orientational correlations may give rise to dynamical heterogeneities and the VFT behavior, we performed the following simulations. \\

\noindent\textbf{Monte Carlo simulations}\\
We performed Monte Carlo (MC) simulations on the 2D $XY$ model in a transverse field.  Its Hamiltonian reads:
\be
H_{\mathrm{XY}}=- J\sum_{i,\mu=\hat{x}, \hat{y}} \cos(\theta_i-\theta_{i+\mu}+F^\mu_i),
\label{xymodel}
\ee
where $\theta_i$ represents the SC phase of the condensate, $J$ the effective Josephson-like interaction between nearest neighboring sites,
and $F_i^\mu$ is the Peierls phase resulting from the minimal substitution prescription, $
F_i^\mu=\frac{2\pi}{\Phi_0}\int_{r_i}^{r_{i+\mu}}A^\mu_i\cdot dr_{\mu}$. The intensity of the magnetic field $B\hat z=\vec{\nabla}\times \vec{A}$ can be expressed in terms of the quantum-flux fraction $f$ passing through a unitary plaquette $f= Ba^2/\Phi_0$, where $a=1$ is the lattice spacing. 

Within the model \eqref{xymodel}, vortices appear as topological excitations of the phase variable $\theta_i$, allowing for the direct characterization~\cite{kato_1993, hattel_flux-lattice_1995, Tanaka_2001, Stroud_1994,Ikeda_2011, Maccari_2021} of the static properties of the solid to liquid transition, via computation of the phase rigidity.  On the other hand, 
dynamical effects have been mainly studied via effective models where vortices are mapped into individual particles~\cite{franz_vortex-lattice_1995, franz_vortex_1994, Zimany_PRB2001, Zimany_PRB2004}. Here we show how model \eqref{xymodel} allows us to address both aspects. Indeed, the solid phase can be identified via the superfluid response, and the dynamical properties of vortices can be characterized by tracking the diffusion of each individual vortex in time at a given temperature.

To establish the superfluid phase, we computed the superfluid stiffness $J_\mathrm{s}$, namely the global phase rigidity of the condensate (see Methods).  At the same time, to establish the orientational order of the VL, we computed  the six-fold orientational order parameter $G_6$, 
\be
\lb{g6}
G_6=\langle \frac{1}{N_\mathrm{v}} \sum_{j=1}^{N} \psi_{6j} \rangle,
\ee
where the sum runs over the $N_\mathrm{v}$ vortices of the lattice, $\psi_{6j}$ indicates the local orientational order relative to the $j-$th vortex (see Methods for details), and $\langle \cdots \rangle$ denotes the thermal average and the average over 10 independent numerical simulations.  The temperature dependence of $J_\mathrm{s}$ and $G_6$ is shown in Fig.\ \ref{Fig:theo}a, along with prototypical images of the corresponding VL; the temperature is expressed in units of $J/k_\mathrm{B}$.  We can identify three distinct phases: a low-$T$ SC phase, where the VL is a pinned solid with a complete hexagonal order ($G_6\simeq 1$); a disordered non-SC  phase at high $T$, where the VL has melted into an isotropic liquid ($G_6\simeq 0$); and an intermediate phase which is a liquid ($J_\mathrm{s}=0$) but with a persistent orientational order of the VL ($G_6 \neq0$) (see also Supplementary information and Fig.~\ref{SK2D}).  The isotropic to hexatic liquid critical temperature $T_{\mathrm{hex}}$ in Fig.~\ref{Fig:theo}a has been identified from the orientational susceptibility $\chi_6$ (see Methods and Fig.~\ref{chi_6}).  Due to the small size of our system we cannot discriminate between a floating solid~\cite{franz_vortex-lattice_1995, hattel_flux-lattice_1995} and a hexatic liquid based on the static properties.  Nonetheless, the dynamic features of the vortices show all the typical fingerprints of the hexatic phase, as they have been identified in 2D soft-colloidal systems~\cite{Rice_2004, Kim_2013, krauth_2015, van_der_meer_dynamical_2015}. To study vortex dynamics we tracked in time each individual vortex, 

\onecolumngrid

\begin{figure}[b!]
    \centering
      \includegraphics[width=\linewidth]{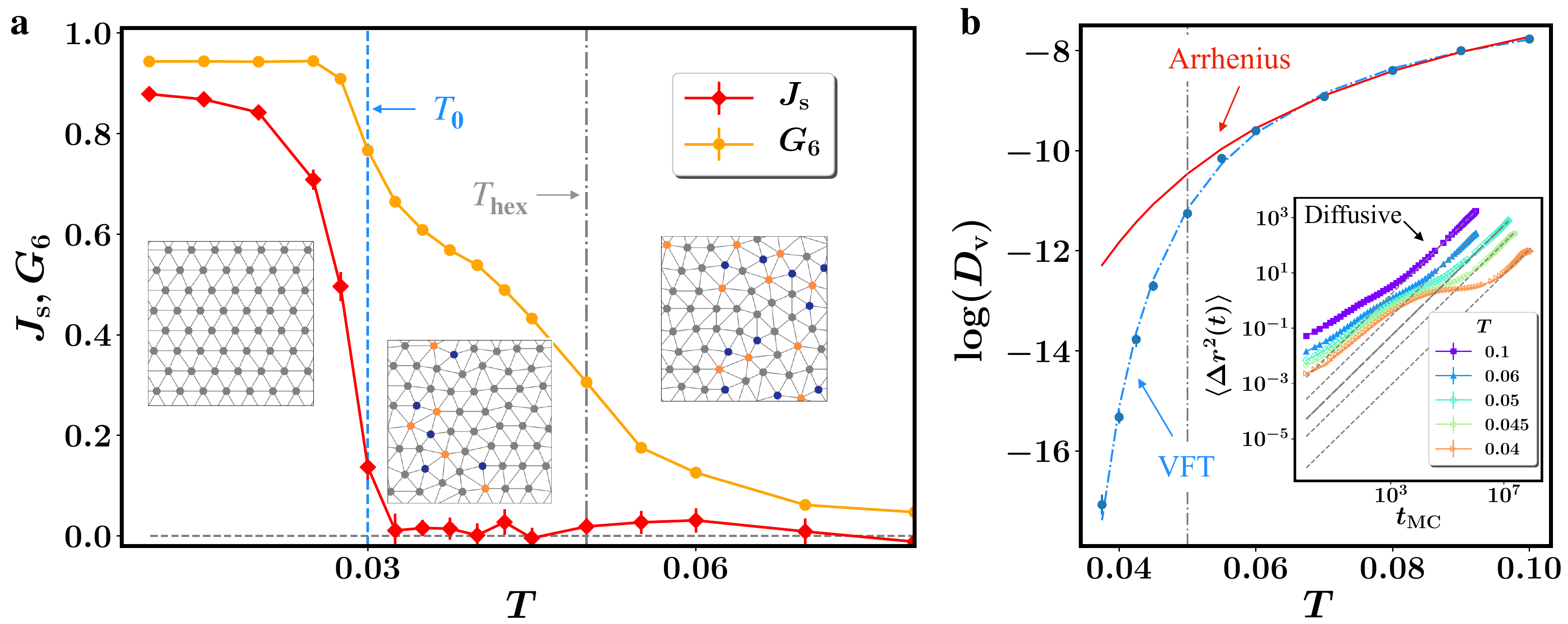}
    \caption{{\bf Numerical simulations of the VL in the 2D XY model.} \textbf{a}, Temperature dependence of the superfluid stiffness $J_\mathrm{s}$ and the $G_6$ orientational order parameter. The insets show snapshots of the VL at temperatures $T=0.02, 0.04, 0.07$ from left to right, respectively.  The blue dashed line marks the temperature $T_0$ where the VFT fit of $D_{\mathrm{v}}$ shown in panel \textbf{b} vanishes, while the gray dot-dashed line indicates the isotropic to the hexatic liquid transition temperature $T_{\mathrm{hex}}$ obtained from the susceptibility of the orientational order parameter (see Methods for more details).  Below $T\simeq T_0$, a pinned solid exists, with true quasi long-range positional and long-range  orientational order, as identified by a finite $J_s$ and $G_6\simeq 1$. Above $T_0$, the quasi long-range positional order is lost due to the appearance of dislocations, formed by bound pairs of disclinations with opposite sign (marked with blue and orange dots in the VL snapshots). As $T$ further increases above $T_\mathrm{hex}$, isolated disclinations appear, leading to a fully isotropic liquid with vanishing orientational order. Notice that, due to the small lattice size, $G_6$ remains finite and disclination pairs always appear nearby.  \textbf{b} Temperature dependence of the vortex diffusion constant $D_\mathrm{v}$, extracted from the diffusive regime of the mean-square displacement shown in the inset.  The dashed lines in the inset mark the linear fits $\langle \Delta r^2(t)\rangle\sim D_\mathrm{v} t$ at selected temperatures.  In the main figure, the dashed blue line is a fit of $D_\mathrm{v}$ with the VFT law \pref{Dv}, while the continuous red line show the Arrhenius fit $D_\mathrm{v, Arrh} =D_\mathrm{v}^0  \exp(-U_\mathrm{v}/T)$.  The vertical gray dot-dashed line indicates $T_{\mathrm{hex}}$. 
    }
\label{Fig:theo} 
\end{figure}
\clearpage 
\twocolumngrid

\noindent and we computed the vortex mean-square displacement  $\langle \Delta r^2(t)\rangle$ at several temperatures, as shown in the inset of Fig.~\ref{Fig:theo}b (also Supplementary information and Fig.~9). 
 
At high $T$, $\langle \Delta r^2(t)\rangle$ directly crosses over from the short-time subdiffusive regime (due to the presence of the numerical square grid) to the typical long-time diffusive behavior, where  $\langle \Delta r^2(t)\rangle\sim D_\mathrm{v } t$, with $D_\mathrm{v}$ the vortex diffusion constant.  However, at the verge of the isotropic- to hexatic-liquid transition,  an additional subdiffusive regime appears with a strong suppression of $\langle \Delta r^2(t)\rangle$ at intermediate time-scales that is the hallmark of dynamical heterogeneities; here this is due to the caging mechanism provided by the finite orientational order in the hexatic phase {(see also Supplementary Video S1).  The signatures of the dynamical heterogeneities persist at longer time scales, with a marked reduction at these temperatures of the asymptotic diffusion coefficient, defined as $D_\mathrm{v} = \frac{1}{4} \lim_{t \to \infty} \langle \Delta r^2(t) \rangle /t$. 
 
The resulting temperature dependence of $D_\mathrm{v}(T)$ is shown in Fig.~\ref{Fig:theo}b.  It is clear that the Arrhenius law (continuous red line) strongly deviates from $D_\mathrm{v}(T)$ at low $T$, in contrast to the VFT law (dashed blue line), which provides a good description of $D_\mathrm{v}(T)$, similar to our experimental findings in the lower-field regime. The most remarkable observation is that the temperature $T_0$ extracted from the VFT fit in Fig.~\ref{Fig:theo}b almost coincides with the temperature where $J_\mathrm{s}$ vanishes (Fig.~\ref{Fig:theo}a), showing an internal consistency between the information available from the vortex dynamics and the static properties of the system in the description of the transition from solid to liquid. \\

\noindent\textbf{Discussion}\\
Our study has revealed similarities between the thermal melting of a 2D VL in weakly pinned $a$-MoGe films and the behavior of fragile glass-forming liquids.  The results of our numerical work strongly support the existence of a glassy dynamics in the absence of strong disorder.  Using the analogy with strong and fragile glasses \cite{CAVAGNA200951}, we can argue that the decrease of $T_0$ with increasing $H$ signals that the apparent glassy state (at $T<T_0$) becomes stronger with $H$. This is consistent with the magnetic field increasing the effective disorder, which further enhances the suppression of vortex diffusivity.   The extrapolation of $T_0(H)$ to zero at $H^{\ast}\simeq 9$~T suggests the existence of a quantum critical point separating the dissipationless solid (for $H<H^{\ast}$) from the SC vortex glass phase that only exists at $T=0$.  Interestingly, a similar field-tuned transition between two superconducting ground states with different ordering of the vortex matter, a vortex solid at lower $H$ and a $T=0$ vortex glass at higher $H$, has been observed also in underdoped copper-oxide high-temperature superconductors \cite{Shi2014, Shi2020}, which are relatively clean quasi-2D systems.  Further insight into the physical mechanisms responsible for the fragile-glass dynamics of the melting of a 2D VL could come from experiments on more disordered thin films and from theoretical exploration of the evolution of the dynamical heterogeneities in the presence of finite disorder.  

\section{Methods}

\noindent\textbf{Samples.} \textit{a}-MoGe films with thickness $t= 22$~nm were grown on surface-oxidized Si substrate through pulsed laser deposition. The reported chemical stoichiometry of these thin films, as seen from the dispersive X-ray analysis, is Mo$_{71\pm1.5}$Ge$_{29\pm1.5}$; their properties have been described in detail elsewhere \cite{Roy_2019,Dutta_2019,dutta2019collective}.  The films were capped with a 2~nm thick Si layer to prevent surface oxidation, and patterned in Hall bar geometry using a shadow mask.  Detailed measurements were performed on two samples with dimensions 0.36~mm (width)$\times$2.8~mm (length), and 1.1~mm distance between the voltage contacts. The two samples exhibited an almost identical behavior. For sample 1, the voltage contact width is 0.05~mm and the zero-field $T_\mathrm{c}= (7.70 \pm 0.05)$~K; for sample 2, the voltage contact width is 0.025~mm and $T_\mathrm{c }=(7.50 \pm0.05)$~K.  $T_\mathrm{c}$ is defined as the temperature at which the resistivity start to rise above the experimental noise floor ($\sim3.6\times10^{-4}$~m$\Omega$cm or $\sim0.5~\Omega$ in resistance).  Gold leads ($\approx50~\mu$m in diameter) were attached to the samples (on top of a Si layer) using the two-component EPO-TEK-E4110 epoxy.  The resulting contact resistances were $\sim 200~\Omega$ each at room temperature.\\

\noindent\textbf{Measurements.}  Resistance was measured using the standard four-probe ac technique ($\sim$13~Hz or 17~Hz) with either SR~7265 lock-in amplifiers or a LS~372 resistance bridge.  Some of the measurements were performed using a dc reversal method with a Keithley 6221 current source and a Keithley 2182A nanovoltmeter.  The excitation current densities were 0.1-10~A\,cm$^{-2}$, depending on the temperature, and low enough to avoid Joule heating.  $dV/dI$ measurements were carried out by applying a dc current bias $I_{\mathrm{dc}}$ and a small ac current excitation ($\sim13$~Hz) through the sample ($I_{ac}=1~\mu$A at $T>1$~K and $I_{ac}=10$~nA at $T<1$~K), while measuring the ac voltage across the sample 
for 150~s and recording the average value for each $I_{\mathrm{dc}}$.

Several cryostats were used to cover a wide range of temperatures and fields: a  HelioxVL $^3$He system ($0.25\leq T\leq 200$~K) with $H$ up to 9 T; a dilution refrigerator ($0.02\leq T\lesssim 1$~K) and a $^3$He system ($0.3 \leq T< 60$~K) in superconducting magnets with $H$ up to 18~T; and a variable-temperature insert ($1.8 < T\leq 200$~K) in a Quantum Design PPMS with $H$ up to 9~T.  The fields, applied perpendicular to the film surface, were swept at constant temperatures.   A low sweep rate of 0.1~T/min was used to avoid heating of the sample due to eddy currents.  Measurements in the dilution refrigerator and the HelioxVL $^3$He system were equipped with filters, consisting of a 1~k$\Omega$ resistor in series with a $\pi$-filter [5 dB (60 dB) EMI reduction at 10 MHz (1 GHz)] in each wire at the room temperature end of the cryostat to reduce high-frequency noise and heating by radiation. Furthermore, the dilution refrigerator and the $^3$He system measurements in 18~T magnets were carried out in the Millikelvin Facility of the National High Magnetic Field Laboratory, which is an electromagnetically shielded room.\\
 
\noindent\textbf{Simulations.}  We have performed Monte Carlo (MC) simulations on a spin system on a square grid with lattice spacing $a=1$, linear size $L=56$ and uniform magnetic field,  whose intensity can be expressed in terms of the quantum-flux fraction $f$ passing through a unitary plaquette $f= Ba^2/\Phi_0$. Here we considered $f= 1/L$, which results in $N_\mathrm{v}=fL^2=56$ vortices with a given vorticity.  Although the system size simulated is smaller than in other numerical simulations of particle systems~\cite{Zimany_PRB2001, Zimany_PRB2004, Rice_2004, Kim_2013, krauth_2015, van_der_meer_dynamical_2015}, it is the state of the art for numerical simulations of vortex lattices within $XY$ models.  Each MC step consists of the local updating of all spins of the lattice by means of the Metropolis-Hastings algorithm. 
All  observables have been computed at equilibrium, as achieved after a certain number $\bar t$ of MC steps. For temperatures above $T_\mathrm{0}$, we identify $\bar t$ with the entrance into the diffusive regime, while below $T_\mathrm{0}$ we estimated that $ \bar{t} \simeq 10^8$ MC steps provides a stable result. 
After discarding 
the first $\bar t$ steps, we reset the Monte Carlo time and proceed with the measurements of both the statical and the dynamical observables.  

 The superfluid stiffness $J_\mathrm{s}$ is defined as the linear response to an infinitesimal phase twist in a given direction, say $\mu$, and it reads:
\be
J_\mathrm{s}^{\mu}= J_d^{\mu} - J_p^{\mu},
\label{js}
\ee
\be
J_d^{\mu}=  \frac{J}{L^2}\left\langle \sum_{i} \cos(\theta_i-\theta_{i+\mu}+F^\mu_i) \right\rangle ,
\label{jd}
\ee
\be
\begin{split}
J_p^{\mu}&=  \frac{J^2}{TL^2} \left\langle \left[\sum_i \sin(\theta_i-\theta_{i+\mu}+F^\mu_i)\right]^2 \right\rangle \\ 
&- \frac{J^2}{TL^2} \left\langle \sum_i \sin(\theta_i-\theta_{i+\mu}+F^\mu_i) \right\rangle^2  .
\end{split}
\label{jp}
\ee
It accounts for two contributions: the diamagnetic part  $J_d^{\mu}$,  proportional to the  energy density of the system, and the paramagnetic part  $J_p^{\mu}$, given by the connected current-current response function.  As in Eq.~\eqref{g6}, here and in what follows, $\langle \dots \rangle$ stands for both the thermal average, performed at equilibrium over all the MC steps, and for the average over ten independent numerical simulations.

The local orientational order parameter $\psi_{6j}$ is obtained by means of a Delaunay triangulation of the VL. It is defined for the hexagonal symmetry, as 
\be
\psi_{6j}= \frac{1}{N_j} \sum_{k=1}^{N_j} e^{6 i \theta_{jk}}, 
\label{psij}
\ee
where $N_j$ is the number of nearest neighbors of the $j$-th vortex, and $\theta_{jk}$ is the angle that the bond connecting the two neighboring vortices $j$ and $k$ forms with respect to a fixed direction in the plane. The global six-fold orientational order parameter $G_6$ is then obtained by summing over all the $N_\mathrm{v}$ vortices and by computing its average, as given by Eq.\ \pref{g6}.  The susceptibility of the orientational oder parameter $\chi_6$ is defined as
\begin{equation}
\chi_6={ \langle \Psi_6^2 \rangle - \langle \Psi_6 \rangle ^2},
\end{equation} 
where $\Psi_6= \frac{1}{N_\mathrm{v}} \sum_{j=1}^{N_\mathrm{v}} \psi_{6j}$ with $ \psi_{6j}$ defined in \eqref{psij}.  The temperature dependence of $\chi_6$ (Fig.~\ref{chi_6}) exhibits a peak at $T=0.05$ that we identify as the critical temperature $T_{\mathrm{hex}}$ between the hexatic and the isotropic liquid phase.

\section{Acknowledgements}
This work was supported by NSF Grant No.~DMR-1707785 and the National High Magnetic Field Laboratory through the NSF Cooperative Agreement No.~DMR-1644779 and the State of Florida.  L.B. acknowledges financial support by MIUR under PRIN 2017 No.~2017Z8TS5B, and by Sapienza University via Grant No. RM11916B56802AFE and RM120172A8CC7CC7. I.M. acknowledges the Carl Trygger foundation through grant number CTS 20:75.  C.D.M. acknowledges support from MIUR-PRIN (Grant No. 2017Z55KCW).

\section{Author contributions}

Thin films were grown and prepared by J.J and S.D.;  B.K.P. and J.T. performed the measurements.  B.K.P. analyzed the data; D.P. and J. T. contributed to the data analysis. I.M. performed Monte Carlo simulations and analyzed numerical results. C.D.M contributed to the data analysis of the numerical results.  J.L., C.D.M., C.C., and L.B. contributed to theory.  I.M, B.K.P., L.B., and D.P. wrote the manuscript.  All authors discussed the results and commented on the manuscript.   D.P., L.B., and P.R.  supervised the project.

\section{Competing Interests}

The authors declare no competing interests.

\onecolumngrid

\clearpage
\section{Supplementary information}

\begin{figure}[h!]
\centering
\includegraphics[width=\linewidth,clip=]{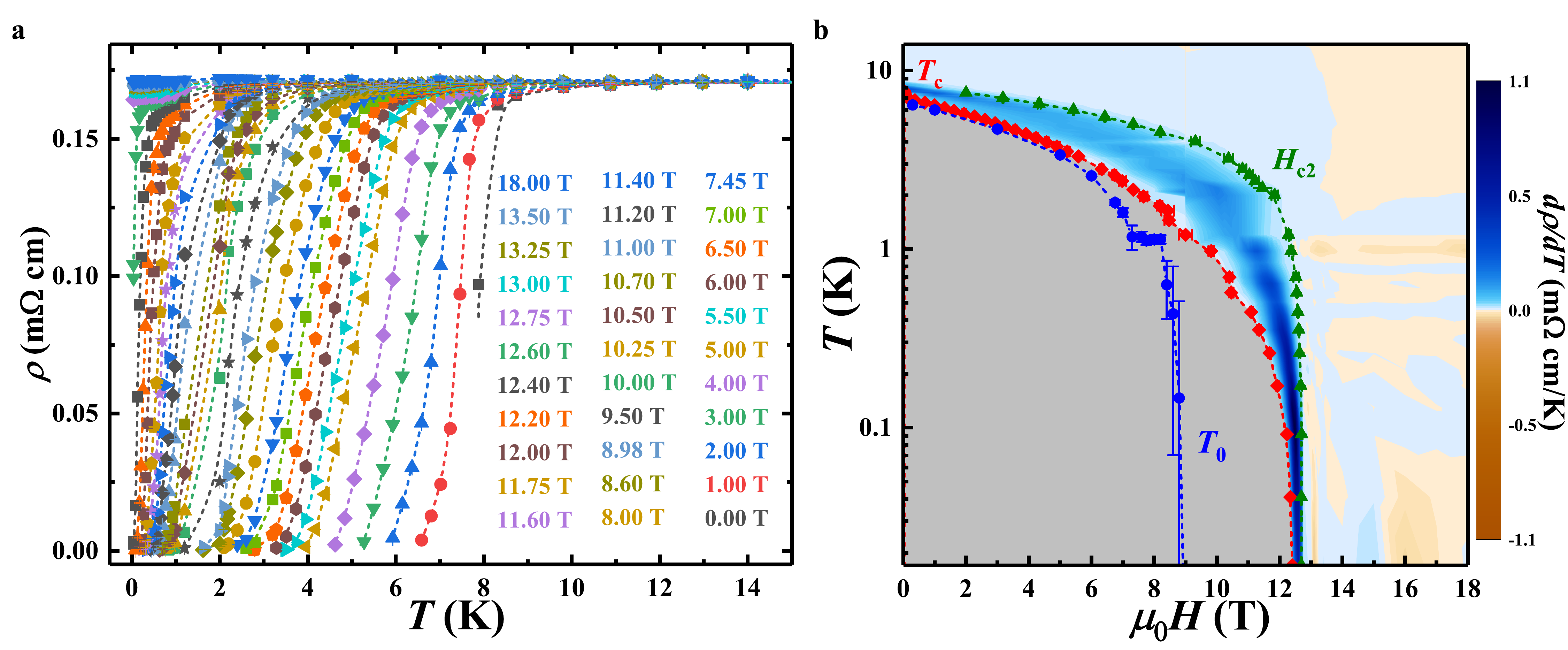}
\caption{{\bf{Resistivity and the $\bm{(T,H)}$ phase diagram of a 22 nm thick \textit{a}-MoGe film.}}  The data are shown for sample 1.  \textbf{a}, $\rho(T)$ at different $H$ over a range of temperatures (0.017 {$\leq$} \textit{T} {$\leq$} 14~K), shown on a linear scale. Dashed lines guide the eye. \textbf{b}, $T$-$H$ phase diagram in a $\log-T$ scale: $T_\mathrm{c}$ (red diamonds), $H_{\mathrm{c}2}$ (green triangles), and $T_0$ (blue dots) represent the superconducting transition temperature, dependent on our experimental resolution, the upper critical field, and the transition to a vortex solid, respectively.  The error bars in $T_0$ reflect the standard error from the VFT fits; error bars in $T_\mathrm{c}(H)$ and $H_{\mathrm{c}2}$ represent the uncertainty in defining the magnetic field where resistance rises above the noise floor or reaches 95\% of the normal state resistance, respectively.  The color map corresponds to $d\rho/dT$.
}
\label{RT-linear}
\end{figure}
\begin{figure}[h!]
\centering
\includegraphics[width=\linewidth]{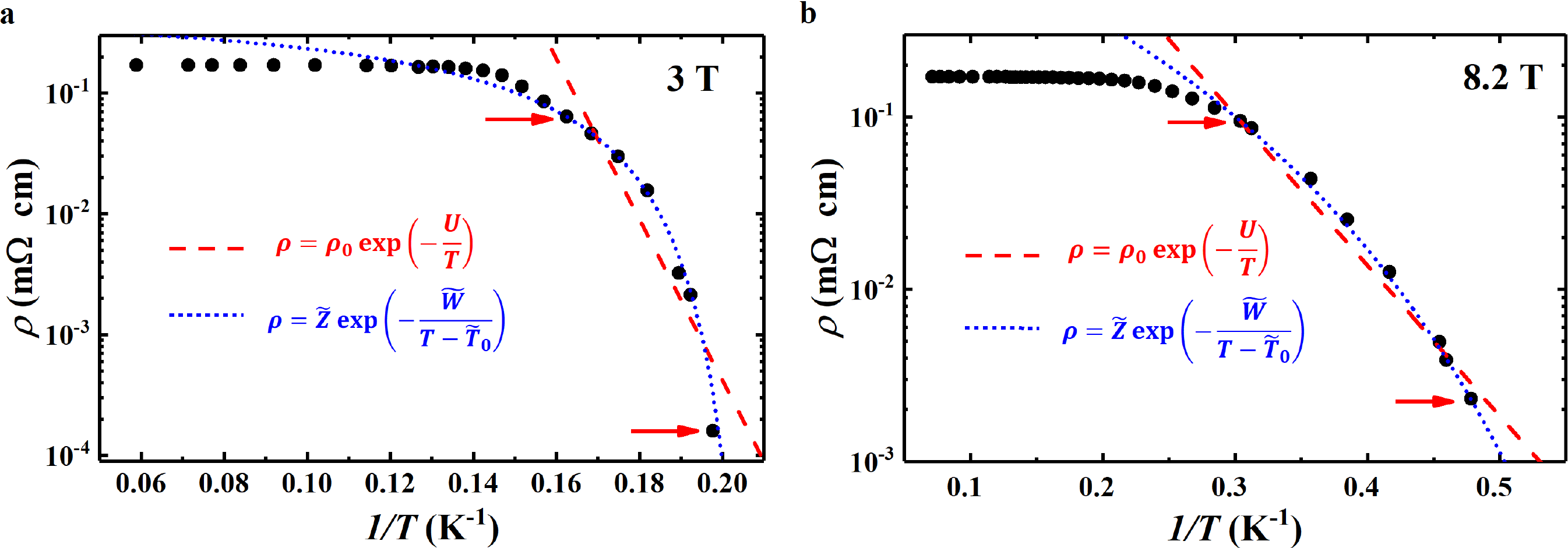}
\caption{ {\bf{Comparison of the Arrhenius and VFT fits of $\bm{\rho(T)}$.}} The data are shown for sample 1.  \textbf{a}, $H=3$~T; $\tilde T_0=(4.71 \pm 0.02)$~K. \textbf{b}, $H=8.2$~T; $\tilde T_0=(1.25 \pm 0.06)$~K.  In both panels, the Arrhenius (red dashed lines) and VFT (blue dotted lines) fits are performed to the data shown within the two red arrows.  The VFT form describes the data better, especially at the lowest $T$.  Since $\tilde T_0$ decreases with increasing $H$, the difference between the two fits also decreases at higher fields.  For the same data, fits to the diffusion coefficient, Eq.~(3), yield $T_0=(4.69 \pm 0.02)$~K and $T_0=(1.14 \pm 0.06)$~K for 3~T and 8.2~T, respectively; these values are the same within error as the corresponding $\tilde T_0$.}
\label{Exp-vs-VFT}
\end{figure}
\begin{figure}[h!]
\centering
\includegraphics[width=0.7\linewidth]{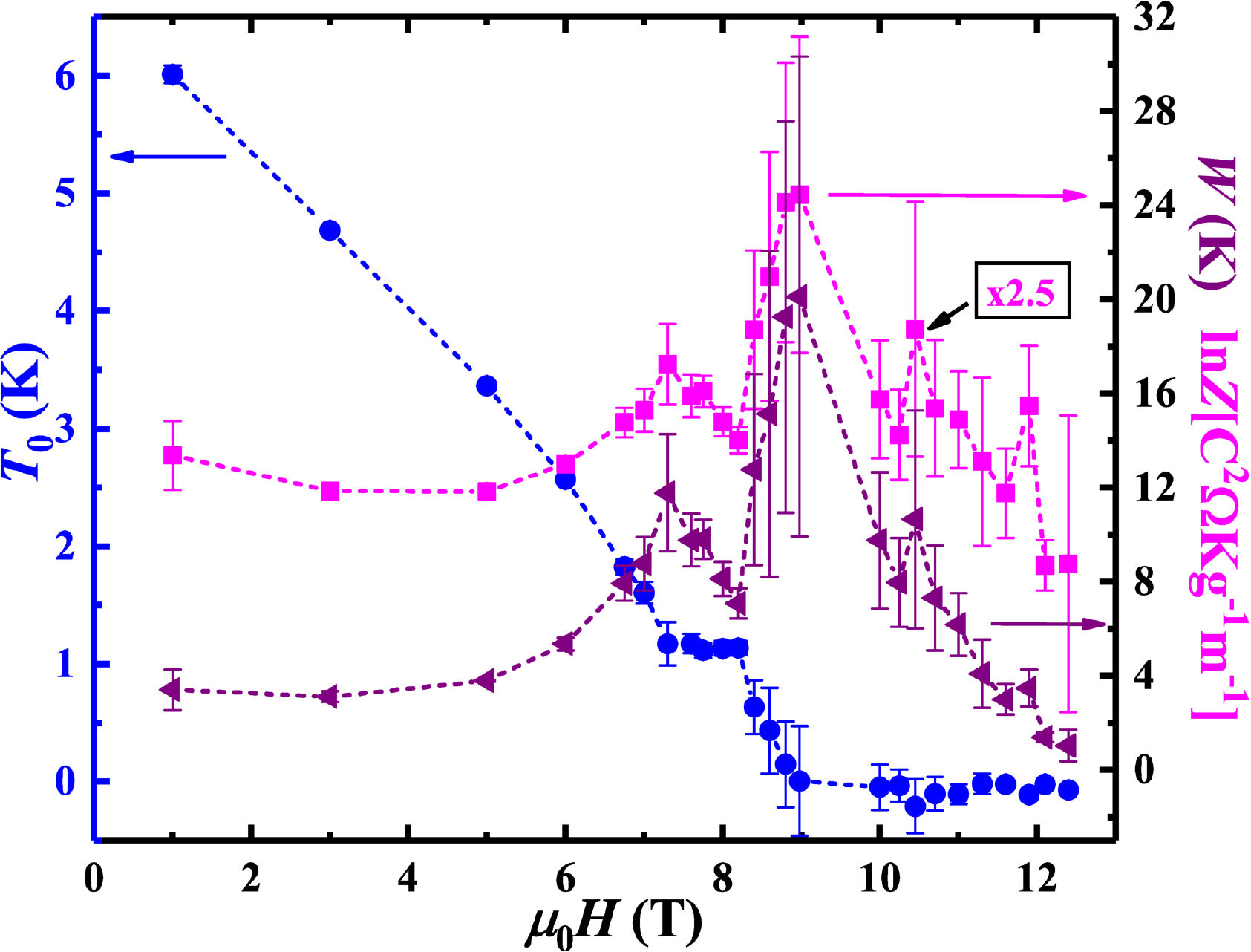}
\caption {{\bf{The VFT fitting parameters $\bm{T_0}$, $\bm{W}$, and $\bm{\ln Z}$.}}  The data are shown for sample 1.  The error bars represent standard errors obtained from the nonlinear fits of ${(\frac{2e}{h})^2k_\mathrm{B}\rho T}$ to the VFT law in Eq.~(3).  A strong correlation between $W$ (purple triangles) and $\ln Z$ (magenta squares) is observed ($\ln Z$ is multiplied by 2.5 on the y-axis for clarity). $T_0$ (blue dots) is independent of the other two parameters.}
\label{VFTFitparadep}
\end{figure}
\begin{figure}[h!]
\centering
\includegraphics[width=0.7\linewidth]{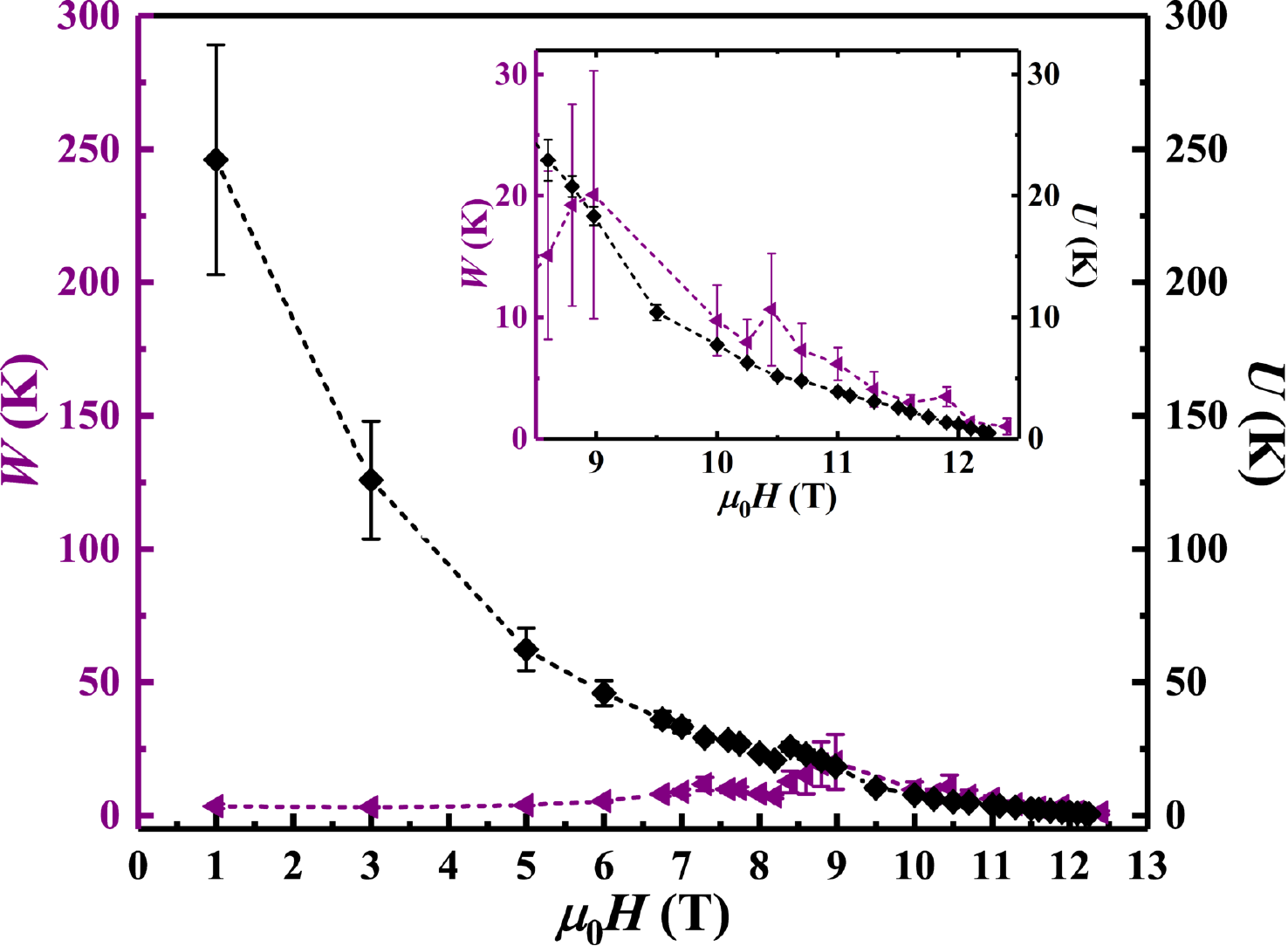}
\caption{ {\bf{Comparison of the parameters $\bm{W}$ (from the VFT fit) and $\bm{U}$ (from the Arrhenius fit).}}  
The data are shown for sample 1.  $W$ (purple triangles) and $U$ (black diamonds) differ considerably for $H<9$~T, but they become equal, within error, for $H \gtrsim 9$~T (see inset) where $T_0=0$, thus confirming the consistency of the analysis.  Dashed lines guide the eye.  The error bars for $W$ and $U$ are standard errors from the nonlinear fit of ${(\frac{2e}{h})^2k_B\rho T}$ to the VFT form and linearized fit of $\rho$ to the Arrhenius form, respectively.}
\label{VFTB-expU}
\end{figure}
\begin{figure}[h!]
\centering
\includegraphics[width=\linewidth]{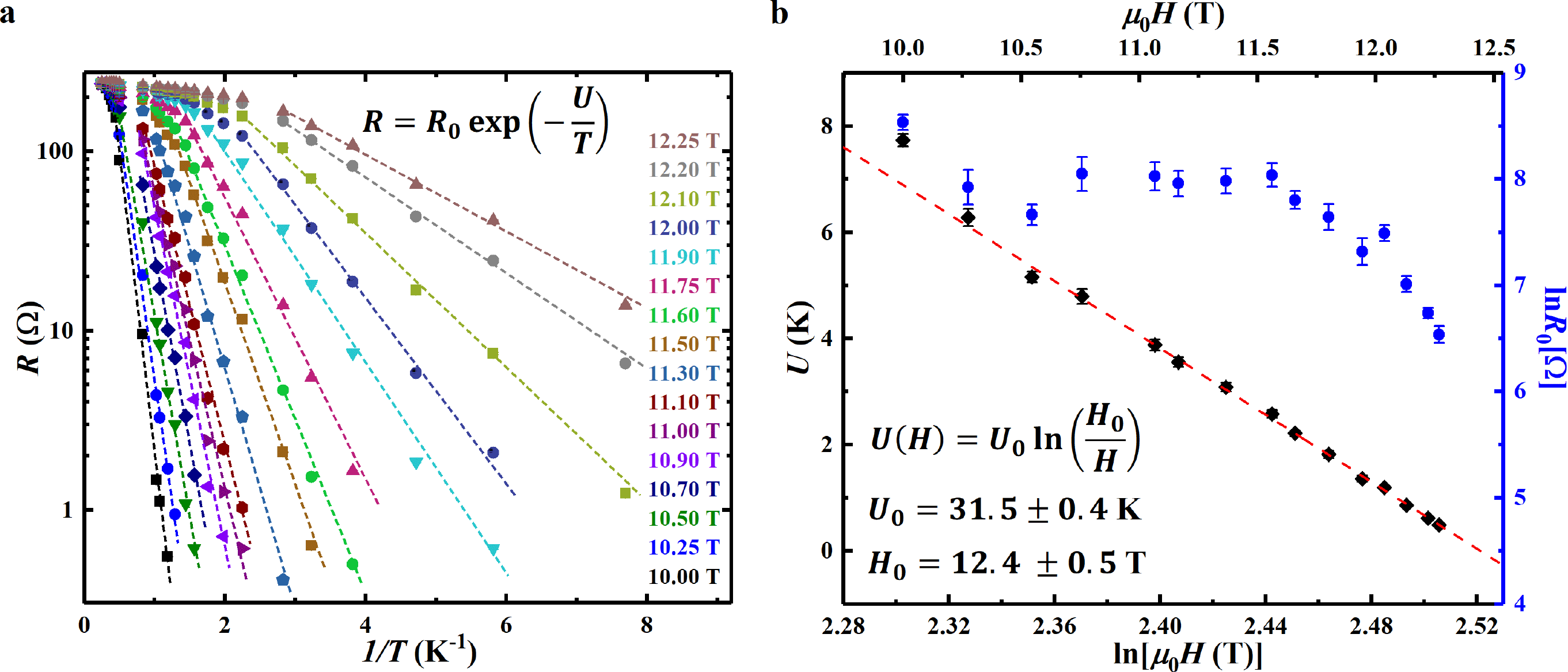}
\caption{{\bf{Arrhenius $\bm{R(T)}$ in the high-field ($\bm{T_0=0}$) regime.}}   The data are shown for sample 1.  \textbf{a}, Resistance $R$ vs $1/T$ for fixed fields.  Dashed lines are linear least-squares fits.  \textbf{b}, The fitting parameters from \textbf{a}, i.e., activation energy $U$ (black diamonds) and $\ln R_0$ (blue dots), vs $H$.  The error bars represent standard errors obtained from the linearized fits in \textbf{a}.  $U(H)=U_{0}\ln(H_0/H)$, where ${U_0}= (31.5 \pm 0.4)$~K and ${H_0}=(12.4 \pm 0.5)$~T.  The red dashed line is a linear least-squares fit.}
\label{Expo-highFields}
\end{figure}
\begin{figure}[h!]
\centering
\includegraphics[width=\linewidth]{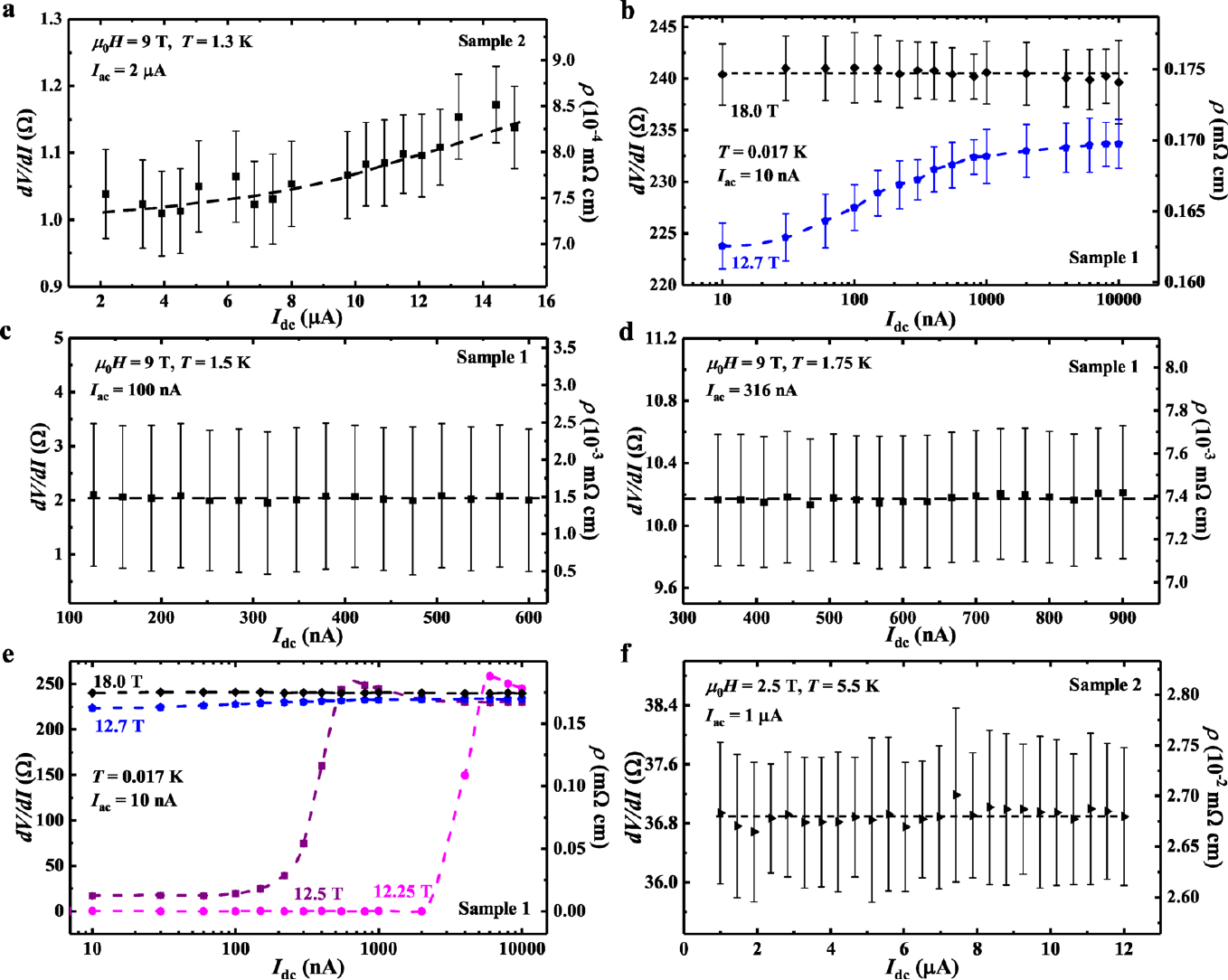}
\caption{{\bf{Differential resistance $\bm{dV/dI}$ as a function of dc current $\bm{I_{\mathrm{dc}}}$ in different regimes.}}  \textbf{a}, In the ``Arrhenius'' region (Fig.~2) at $T=1.3$~K and $H=9$~T, a strong nonlinearity is observed, consistent with the motion of vortices in the presence of disorder; $T_{\mathrm{c}}(H=9$~T$)=1.1\pm 0.2$~K.  \textbf{b}, Similar nonlinear $dV/dI$ vs $I_{\mathrm{dc}}$ in the ``Arrhenius'' region for $H=12.7$~T and $T=0.017$~K.  At a much higher field ($H=18.0$~T$>H_{\mathrm{c}2}$), in the normal state, $dV/dI$ vs $I_{\mathrm{dc}}$ is ohmic.  The nonlinear behavior becomes unobservable also as $T$ increases, as shown in \textbf{c}, for $H=9$~T at $T=1.5$~K, and \textbf{d}, for $H=9$~T at $T=1.75$~K, at lower currents than in \textbf{a}.  \textbf{e}, The evolution of $dV/dI$ vs $I_{\mathrm{dc}}$ with increasing $H$, as shown, at $T=0.017$~K.  The data for $H=18.0$~T and $H=12.7$~T are the same as in \textbf{d}.  When $\rho$ drops below the noise floor, e.g. for $H=12.25$~T, a finite $I_{\mathrm{dc}}\simeq 2~\mu$A, i.e. a critical current, is needed to depin a measurable number of vortices.   At somewhat higher $H=12.5$~T, $\rho$ is nonzero at low current bias, consistent with the vortex creep due to thermal fluctuations.  The depinning effect becomes observable at the critical current $I_{\mathrm{dc}}\sim 0.1~\mu$A, which is lower than that for $H=12.25$~T, as expected.  \textbf{f}, $dV/dI$ vs $I_{\mathrm{dc}}$ is ohmic at $T\approx 5.5$~K and $H=2.5$~T, where $\rho(T)$ obeys the VFT law.  The error bars in all panels correspond to $\pm1$ SD obtained from averaging the ac voltage over 150~s at a fixed $I_{\mathrm{dc}}$.  Dashed lines guide the eye.  The $T$ fluctuations for these measurements were less than 5~mK.}
\label{dVdI}
\end{figure}
\clearpage
\section{S1. Monte Carlo simulations: {Static characterization of the vortex lattice} }

To further characterize the three phases found, we have computed the structure factor of the vortex lattice defined as:
\be
S(\mathbf{k})= \frac{1}{2 N_\mathrm{v}^2} \sum _{i, j} \exp\Big[ i \mathbf{k} \cdot ( \mathbf{r_{i} - r_{j}}) \Big]  \left\langle \nu( \mathbf{r_{i}}) \nu(\mathbf{r_{j}}) \right\rangle ,
\ee
where $N_\mathrm{v}$ is the total number of vortices, $\nu(\mathbf{r_i})$ is the local vortex density, equal to one if a vortex occupies the site $\mathbf{r_i}$ and zero otherwise, and $\mathbf{k}$ is the vortex lattice reciprocal vector. In Fig.~\ref{SK2D}, we show $S(\mathbf{k})$ computed at three different temperatures: $T=0.02, 0.04, 0.07$, corresponding respectively to the solid, hexatic-liquid, and isotropic-liquid phase.  At high temperature, the structure factor presents the typical circular symmetry of an isotropic liquid. Decreasing the temperature, such symmetry breaks down and the Bragg-peaks structure appears showing six well-defined spots. Finally, in the solid state the Bragg peaks become well-defined even at large $k$. Let us highlight that the structure-factor anisotropy observed at low temperature is due to the commensurability of the VL with respect to the underlying square lattice of spins. The VL has indeed two ways to align with respect to the underlying square lattice: either to be perfectly commensurate with the $\hat{x}$-axis or with the $\hat{y}$-axis.
\begin{figure}[h!]
\centering
\includegraphics[width=\linewidth]{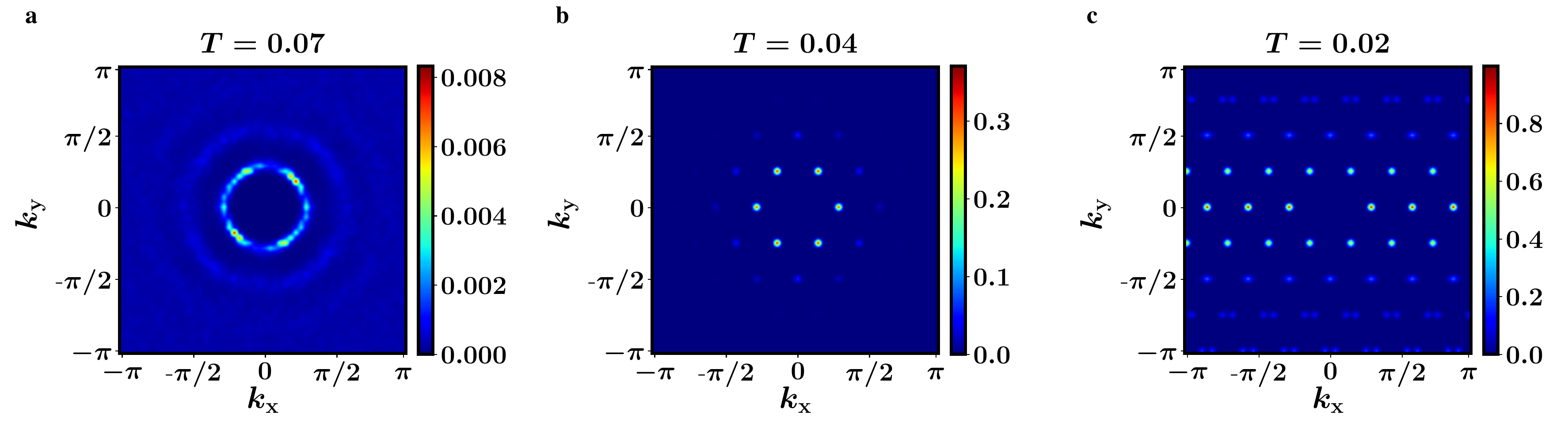}
\caption{{\bf{Structure factor of the vortex lattice in the three phases.}} Structure factor computed for a given sample at three different temperatures corresponding to the three different phases: $T=0.07$ isotropic liquid phase;  $T=0.04$ hexatic liquid phase; $T=0.02$ solid phase. To highlight the main features of the structure factor we have fixed  $S(\mathbf{k}=0)=0$.}
\label{SK2D}
\end{figure}

In order to identify the isotropic to hexatic liquid transition, we have computed the orientational susceptibility $\chi_6$.  The susceptibility of the orientational oder parameter $\chi_6$ is defined as
\begin{equation}
\chi_6={ \langle \Psi_6^2 \rangle - \langle \Psi_6 \rangle ^2},
\end{equation} 
where $\Psi_6= \frac{1}{N_\mathrm{v}} \sum_{j=1}^{N_\mathrm{v}} \psi_{6j}$ with $ \psi_{6j}$ defined in Eq.~(9) of the main text.

The temperature dependence of $\chi_6$ (Fig.~\ref{chi_6}) exhibits a peak at $T=0.05$ that we identify as the critical temperature $T_{\mathrm{hex}}$ between the hexatic and the isotropic liquid phase.
\begin{figure}[h!]
\centering
\includegraphics[width=0.5\linewidth]{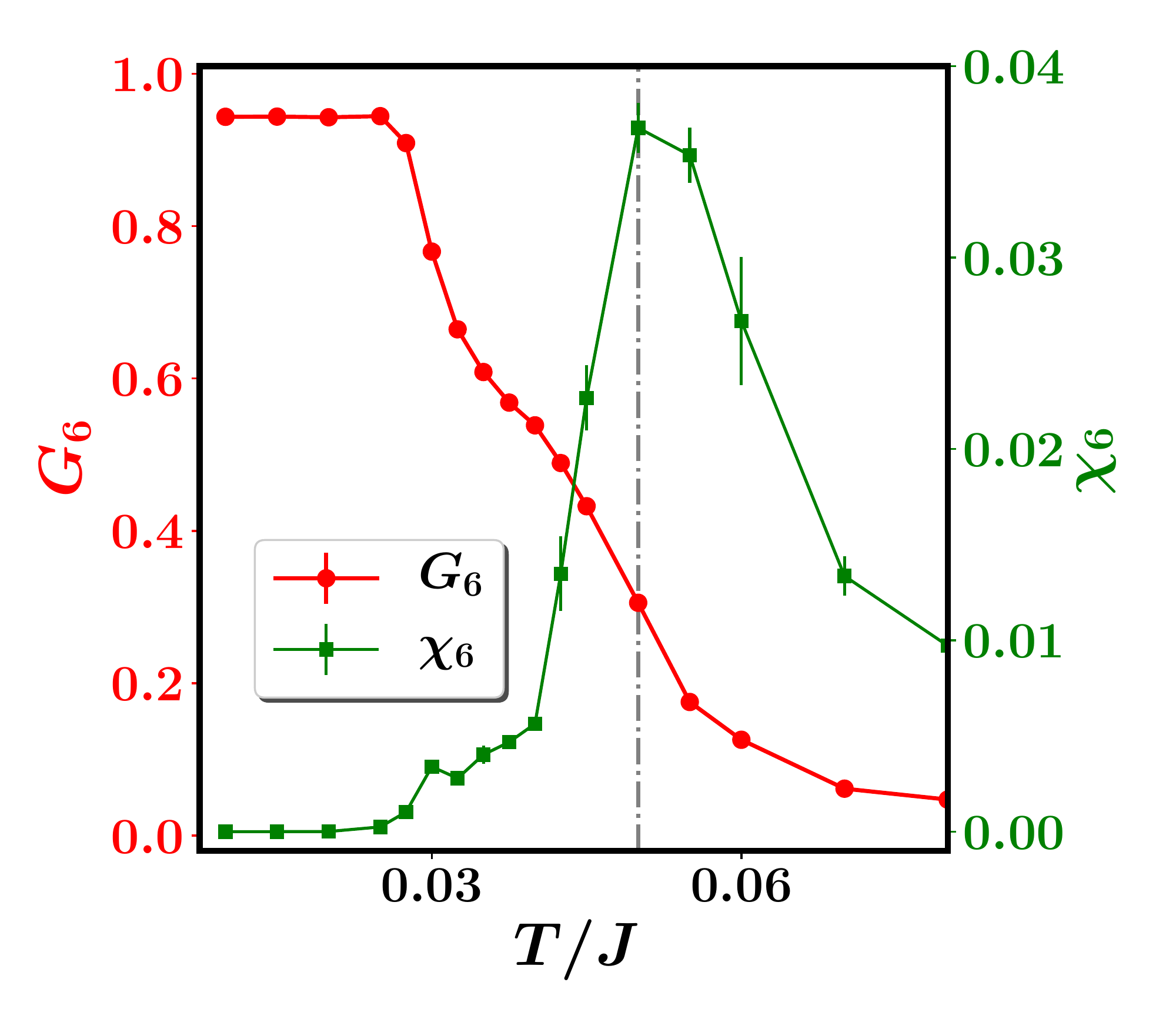}
\caption{{\bf{Orientational order parameter and susceptibility.}} Orientational order parameter $G_6$ and the corresponding orientational susceptibility $\chi_6$ as function of the temperature in units of the coupling constant of the XY model $J$. The peak of the orientational susceptibility identifies the temperature $T_{\mathrm{hex}}$ separating the hexatic from the isotropic liquid phase. }
\label{chi_6}
\end{figure}

\section{S2. Monte Carlo simulations: signatures of heterogeneous dynamics}

The heterogeneous nature of the vortex dynamics in the hexatic phase has its fingerprints in different observables.  Together with the mean-square displacement $\langle \Delta r^2(t) \rangle$, reported in the main text, we also computed the self-part of the intermediate scattering function $F_s(|\mathbf{k^*}|, t)$, and the non-Gaussian parameter $\alpha_2(t)$.  The resulting trends in time for different temperatures are shown in Fig.~\ref{Dynamic}, where the time variable $t$ labels the discrete MC steps.  To highlight the onset of the hexatic phase, the curves at the verge of the isotropic-liquid to the hexatic-liquid transition ($T=0.05$) have been plotted in gray.  

The self-part of the intermediate scattering function is a dynamical autocorrelation function for the VL and it is defined as:
\be
F_s(|\mathbf{k}^*|, t) = \frac{1}{N_\mathrm{v}} \sum_{j=1}^{N_\mathrm{v}} \frac{1}{(t_{M} - t)} \sum_{t_0 =0}^{t_{\mathrm{M}}-t} \frac{1}{N_{\mathbf{k}}} \sum_{\mathbf{k} : |\mathbf{k}| = |\mathbf{k}^*|} \overline{ \exp\{i \mathbf{k} [ \mathbf{r}_j(t_0 +t) - \mathbf{r}_j(t_0)] },
\label{Self}
\ee 
where $|\mathbf{k^*}|= 2\pi/a_\mathrm{v}$ (with $a_\mathrm{v}$ the lattice spacing of the VL) is the reciprocal vector at which the structure factor shows its first peak (see Fig.~\ref{SK2D}), $ \mathbf{r}_j(t)$ is the position of the j-$th$ vortex at the MC time $t$ and $t_{\mathrm{M}}$ is the largest MC time used in the simulations. The thermal average is thus the sum over all the possible $t_0$ for a given time $t$, while $\overline{(\dots)}$ stands for the average over ten independent simulations.  
\begin{figure}[b!]
\centering
\includegraphics[width=\linewidth]{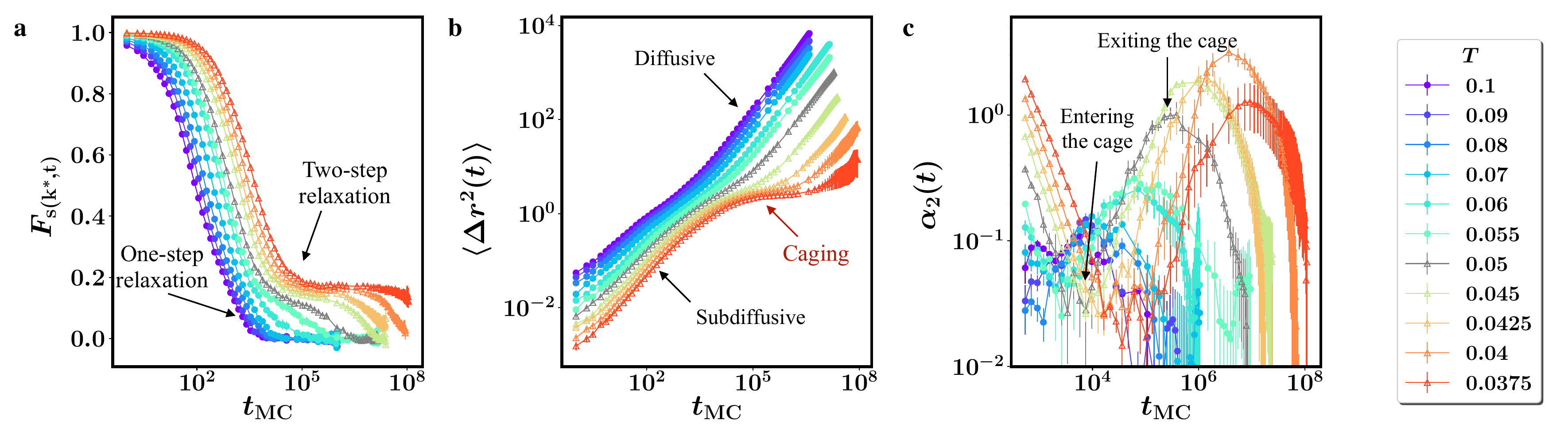}
\caption{{\bf{Dynamic characteristics of the liquid phase.}}
\textbf{a}, The self-part of the intermediate scattering function computed at  $|\mathbf{k^*}|= 2\pi/a_\mathrm{v}$, with $a_\mathrm{v}$ the lattice spacing of the VL, \textbf{b}, the mean-square displacement, and \textbf{c}, the non-Gaussian parameter as function of the MC time for different temperatures. The two arrows, which indicate entry into and exit from the cage, refer to the temperature $T=0.05$ plotted in gray. In all panels, gray lines correspond to the hexatic critical temperature $T_{\mathrm{hex}}=0.05$.}
\label{Dynamic}
\end{figure}
From Fig.~\ref{Dynamic}a, one can see that, in the isotropic liquid phase, $F_s(|\mathbf{k}_{max}|, t)$ decays exponentially to zero with a unique relaxation time. On the other hand, as the hexatic phase is approached, it starts showing a plateau, which increases with decreasing $T$. This  two-step relaxation decay is another typical signature of a caging mechanism~\cite{CAVAGNA200951} due in this case to the onset of the hexatic phase.  The mean-square displacement, already shown in the inset of Fig.~3b,  is defined as:
\be
\langle \Delta r^2(t) \rangle = \frac{1}{N_\mathrm{v}} \sum_{j=1}^{N_\mathrm{v}}  \frac{1}{(t_{\mathrm{M}} - t)} \sum_{t_0=0}^{t_{\mathrm{M}}-t}  \overline{\left|\mathbf{r}_j(t_0 +t) - \mathbf{r}_j(t_0) \right|^2 }.
\label{msd}
\ee
As discussed in the main text, at the onset of the hexatic liquid phase an intermediate subdiffusive regime appears (Fig.~\ref{Dynamic}b).  This signals an inhibition of the particle motion due to the cage formed by the neighboring particles, similarly to what happens in supercooled liquid and glassy systems. 

The intermediate plateau observed both in the mean-square displacement and in the intermediate scattering function signals the emergence of a heterogeneous dynamics within a certain time scale. To further investigate this behavior, we have computed the non-Gaussian parameter
\be
\alpha_2(t)= \frac{1}{2} \frac{\left \langle \Delta \mathbf{r}^4(t) \right\rangle }{\left \langle \Delta \mathbf{r}^2(t) \right\rangle^2 } - 1, 
\ee
which quantifies the heterogeneity of the dynamics in terms of strength and extent in time \cite{Raman_1964}.  We find (Fig.~\ref{Dynamic}c) that, at very short times, the VL displays a strongly heterogeneous dynamics for all the temperatures analyzed. This anomalous behavior is due to the underlying squared grid of spins that forces vortices to move only along four possible directions: $\pm \hat{x}$ and $\pm \hat{y}$. With increasing time, the direction of motion becomes less sensitive to the underlying grid and $\alpha_2(t)$ decreases. Apart from the initial heterogeneity, at high temperatures the VL shows a homogenous dynamics. On the other hand, at the onset of  the hexatic phase, the non-Gaussian parameter starts displaying a dome at longer time scales. This additional signature of heterogeneity is another indication of a caging mechanism triggered by the onset of the hexatic phase. With decreasing temperature the height of the dome increases, signalling the increase of the heterogeneous dynamics strength. At the same time, the peak of the dome moves to longer times, reflecting the increase of the time scale at which vortices escape from their cage. 

\section{S3. Brief description of the supplementary video}

\emph{Supplementary Video S1: Heterogeneous vortex motion in the hexatic phase.} Images of the vortex motion at $T=0.045$ for different MC times. 
In this regime, the dynamics is strongly heterogeneous in both space and time. The vortices remain indeed trapped for a long time in the cage formed by their neighbors before exiting in a collective burst along the symmetry axis of the vortex lattice. 

In the video, the vortices are represented as colored disks, centered at a given position ${\bf x_i}$ at time $t_i$. Each frame shows  vortex evolution over ten consecutive discrete time steps $t_i$. To help visualize the time evolution, we assigned to each disk a radius $r_i$ that is increasing with increasing time $t_i$. As a consequence, larger disks identify the vortex position at larger times.  In addition, we add a solid black line connecting the centers of the disks to help visualize the vortex motion as a function of time. The grey lines in the background show the trails left by the vortices during the whole simulation. The horizontal line at the bottom of the image indicates the time flow.  Note that the time steps $t_i$ are not equally spaced with respect to the MC time. Indeed, for the sake of the memory allocation, we have stored the data each $t_{\mathrm{MC}}= int(A^q) + kA^{N_\mathrm{q}} $, with $q \in [0, N_\mathrm{q}]$, $k \in [0, N_\mathrm{k}] $ and $A=1.3$.

\end{document}